\documentclass[12pt, onecolumn,a4paper,draftclsnofoot]{IEEEtran}
\usepackage{amssymb,amsmath}
\usepackage{cite}
\usepackage{graphicx,subfigure}
\usepackage{psfrag}
\usepackage{url}
\usepackage[latin1]{inputenc}

\usepackage[absolute,overlay]{textpos}
\usepackage{tikz}
\usetikzlibrary{arrows,calc,decorations.markings}
\usetikzlibrary{topaths}
\usetikzlibrary{shapes,trees,snakes}
\usetikzlibrary{arrows}
\usetikzlibrary{shadows}
\usetikzlibrary{positioning}
\usetikzlibrary{matrix}
\usetikzlibrary{shapes.geometric}
\usetikzlibrary{decorations.pathmorphing}
\usepgflibrary{patterns}
\usetikzlibrary{calc}
\usetikzlibrary{fit}					
\usetikzlibrary{backgrounds}	

\usepackage{pgf}
\usetikzlibrary{arrows,automata}
\usepackage[latin1]{inputenc}

\begin{document}

\title{Ultra Reliable Short Message Relaying with Wireless Power Transfer}
\author{
	\IEEEauthorblockN{	Onel L. Alcaraz López\IEEEauthorrefmark{1}, %
						Hirley Alves\IEEEauthorrefmark{2}, 
						Richard Demo Souza\IEEEauthorrefmark{3}, and
						Evelio Martín García Fernández\IEEEauthorrefmark{1}
					}\\
	\IEEEauthorblockA{\IEEEauthorrefmark{1}Federal University of Paraná (UFPR), Curitiba, Brazil\\
					}
					\IEEEauthorblockA{\IEEEauthorrefmark{2}Centre for Wireless 	Communications (CWC), Oulu, Finland\\}
					\IEEEauthorblockA{\IEEEauthorrefmark{3}Federal University of Technology - Paraná (UTFPR), Curitiba, Brazil\\}
						onel2428@gmail.com, %
					 	hirley.alves@oulu.fi,
					 	richard@utfpr.edu.br,
					 	evelio@ufpr.br}
\maketitle

\begin{abstract}

We consider a dual-hop wireless network where an energy constrained relay node first harvests energy through the received radio-frequency signal from the source, and then uses the harvested energy to forward the source's information to the destination node. The throughput and delay metrics are investigated for a decode-and-forward relaying mechanism at finite blocklength regime and delay-limited transmission mode. We consider ultra-reliable communication scenarios under discussion for the next fifth-generation of wireless systems, with error and latency constraints. The impact on these metrics of the blocklength, information bits, and relay position is investigated.

\end{abstract}

\section{Introduction}
Energy Harvesting (EH) techniques have recently drawn significant attention as a potential solution to prolonging the lifetime of energy constrained wireless devices \cite{Ozel.2011}. 
Amongst those, the Wireless Power Transfer (WPT) has recently emerged as an attractive solution to power nodes in future wireless networks \cite{Deniz.2014}, \cite{Visser.2013}. This comes with the implicit advantage that radio-frequency (RF) signals can carry both energy and information, which enables energy constrained nodes to scavenge energy and receive information \cite{Varshney.2008}, \cite{Grover.2010}. Nevertheless, the current state-of-the-art in electronic circuits makes impossible to harvest the received energy after passing through an information decoder \cite{Huang.2013}. This motivated the design of practically realizable receivers which separate information decoding and EH processes. The most common used techniques to do so are based on time-switching or power splitting \cite{Zhou.2013}. Based on these, authors in \cite{Nasir.2013} propose two WPT protocols called Time Switching-based Relaying (TSR) and Power Splitting-based Relaying (PSR) to be implemented by an energy constrained relay node in order to assist a single source-destination communication link. These protocols are then evaluated in terms of throughput under the delay-limited and delay-tolerant transmission modes, and for Amplify-and-Forward (AF) relaying. The implementation of the TSR protocol is analyzed later in \cite{Nasir.2015} for Decode-and-Forward (DF) relaying and a fixed preset relay transmit power.
All the above studies are under the ideal assumption of communicating with blocks of infinite length, using the  performance metrics based on Shannon's channel capacity and its extension to nonergodic channels$-$the outage capacity. In spite of their asymptotic nature, they have proven to be useful to design current wireless systems in which delay constraints are typically above 10 ms, which allows the use of sufficiently long packets \cite{Devassy.2014}. However, when the transmitted packets are short, these assumptions do not fit properly. Indeed, short packets are essential to support Ultra-Reliable Communication (URC) \cite{Popovski.2014}, which is a novel operation mode under discussion for the next fifth-generation (5G) of wireless systems. URC with short packets (URC-S: URC over a Short Term) focus on how to deliver a certain portion of data under a very stringent latency requirement, e.g. 2ms, and a given target error probability, e.g. $10^{-5}$, \cite{Durisi.2014}. This will be the typical form of traffic generated by sensors, critical connections for industrial automation and reliable wireless coordination among vehicles \cite{Popovski.2014,Johansson.2015,Yilmaz.2015}. 
Under these scenarios, the fundamental performance limit is the maximum achievable rate $R^*(n,\epsilon)$ at a given block length $n$ and error probability $\epsilon$. This metric has been recently characterized in \cite{Polyanskiy.2010,Yang.2013} for both Additive White Gaussian Noise (AWGN) and fading channels.
On the other hand, scenarios with rate adaptation are analyzed in \cite{Schiessl.2016} for imperfect Channel State Information (CSI) at the transmitter.  
Work in \cite{Hu2016} focuses on the performance of multi-terminal wireless industrial networks with stringent low-latency requirements, investigating relay selection among the participating terminals and best antenna selection at the access point of the network. In both schemes, they incorporate the cost of acquiring instantaneous CSI at the access point within the transmission deadline, and analyze the impact of a target error probability.

Only recently, scenarios with EH have been analyzed at finite blocklength regime \cite{Yang.2014,Shenoy.2016,Polyanskiy.2016,Tandon.2016,Khan.2016}. In \cite{Shenoy.2016} the authors consider an AWGN channel where the transmitter is powered by EH only. Then, a lower bound on the maximal codebook at finite code lengths is provided and shown that it improves upon previously known bounds. In \cite{Polyanskiy.2016}, the achievable channel coding rate and mean delay of a point-to-point EH wireless communication system with finite blocklength is investigated for the AWGN channel. All these works are abstracted from the EH process, which could be practical for systems where the EH techniques are not based on the RF signals. Meanwhile, further analyzes are required when the energy harvested comes from a WPT process. Accordingly, the authors of \cite{Tandon.2016} consider a receiver that uses the same received signal both for decoding information and for harvesting energy through the power splitting technique. Then, they study subblock energy-constrained codes (SECCs) and provide a sufficient condition on the subblock length to avoid power outage at the receiver. The work in \cite{Khan.2016} analyzes a system where a node, charged by a power beacon through a Rayleigh channel, attempts to communicate with a receiver over a noisy channel. The system performance is investigated when varying key parameters such as the number of channel uses for PT and those for information transfer. Conversely, the present paper aims at assessing URC scenarios with a target error probability and latency constraints. We analyze, at finite blocklength regime and delay-limited transmission mode, the TSR protocol studied in \cite{Nasir.2013,Nasir.2015}. The system is a dual-hop communication network with Nakagami-m fading, where the DF relay first harvests energy from the source transmission and then uses that energy to forward the received signal to the destination node.  We derive an analytical approximation for the throughput at these scenarios and validate its accuracy through simulations. Results show the trade-off between reliable communications and latency constraints. While the first requires increasing the number of channel uses for reliable PT and/or information transfer, the second imposes a restriction on this number. The impact of the required channel uses and message length on the system performance is investigated in terms of throughput and delay. The optimum attainable throughput is shown to be independent of the message length. However, the smaller this number, less channel uses are required, improving the delay. Moreover, in a fixed delay scenario, when the message length decreases the optimum number of channel uses for PT increases. Finally, the optimal relay position is shown to be closest to the source, which differs from traditional relaying without WPT.

The paper is organized as follows. Section \ref{system} presents the system model and assumptions. Section \ref{DF} discusses the throughput and delay metrics for DF relaying at finite blocklength. Section \ref{results} presents the numerical results. Finally, Section \ref{conclusions} concludes the paper.

\section{System Model and Assumptions}\label{system}

We consider a dual-hop cooperative network where an energy constrained relay node ($\mathbb{R}$) assists the transmissions from the source ($\mathbb{S}$) to the destination ($\mathbb{D}$).  We assume that there is no direct link between $\mathbb{S}$ and $\mathbb{D}$, which is meaningful in many real-world scenarios \cite{Fouladgar.2012,Krikidis.2012,Chen.2014,Ding.2014}, and thus the system performance relies fundamentally on $\mathbb{R}$. Nakagami-m quasi-static block-fading channels are assumed, where fading process is considered to be constant over the transmission of a block and independent and identically distributed from block to block. Normalized channel gains from $\mathbb{S}$ to $\mathbb{R}$ and from $\mathbb{R}$ to $\mathbb{D}$ are denoted by $\tilde{h}$ and $\tilde{g}$, respectively. Let $h=|\tilde{h}|^2$ and $g=|\tilde{g}|^2$, then $h,g\sim\Gamma(m,1/m)$ with shape factor $m$ and Probability Density Function (PDF) $f_z(z|m)=\frac{m^m}{\Gamma(m)}z^{m-1}e^{-mz}$ where $z\in\{h,g\}$. Additionally, the distances from $\mathbb{S}$ to $\mathbb{R}$ and from $\mathbb{R}$ to $\mathbb{D}$ are denoted by $d_1$ and $d_2$, respectively.

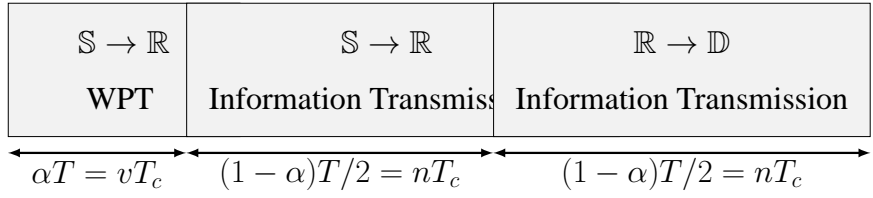
\begin{figure}[!t]
	\centering
	\resizebox{120mm}{!}{ \def\x{{\mathbf x}}
\def\L{{\cal L}}
\pgfarrowsdeclarecombine{|<}{>|}{|}{|}{latex}{latex}
\def\Dimline[#1][#2][#3]{
	\begin{scope}[>=latex] 
		\draw[<->,very thick,
		decoration={markings, 
			mark=at position .5 with {\node[black] at (0,-0.4) {\Large{#3}};},
		},
		postaction=decorate] #1 -- #2 ;
	\end{scope}
}

\begin{tikzpicture}
{\Large
		\centering
		\foreach \c/\i [count=\n] in
		{gray!10/{\begin{tabular}{c}
                 ~~~~~~$\mathbb{S}\rightarrow\mathbb{R}$~~~~~~ \\
                 WPT~
				\end{tabular}},gray!10/{\begin{tabular}{c}
				$\mathbb{S}\rightarrow\mathbb{R}$~~~~~\\
				Information Transmission~~~~~~~
				\end{tabular}
				 },gray!10/{\begin{tabular}{c}
				 $\mathbb{R}\rightarrow \mathbb{D}$\\
			     Information Transmission
				 \end{tabular}
				 }}
		\node[draw,fill=\c,minimum height=1.8 cm,minimum width = 1 cm,xshift=\n*5.2cm] (N\n){\i};
}	
		
		
{\Large
		\Dimline[($(N1.south west)-(0,0.3)$)][($(N2.south west)-(0,0.3)$)][$\alpha T=vT_c$];
		\Dimline[($(N2.south west)-(0,0.3)$)][($(N3.south west)-(0,0.3)$)][$(1-\alpha)T/2=nT_c$];
		\Dimline[($(N3.south west)-(0,0.3)$)][($(N3.south east)-(0,0.3)$)][$(1-\alpha)T/2=nT_c$];
	}
\end{tikzpicture}}
	\caption{TSR protocol for EH and information processing at $\mathbb{R}$ with finite blocklength. This figure is adapted from \cite[Fig.~2a]{Nasir.2013}.}\label{fig:TSR}
	\vspace{-1mm}
\end{figure}

The energy required by $\mathbb{R}$ is obtained from a WPT process according to the TSR protocol \cite{Nasir.2013} as shown in Fig.~\ref{fig:TSR}. 
Node $\mathbb{S}$ uses first $v$ channel uses to  power $\mathbb{R}$. Then, during $n$ consecutive channel uses, $\mathbb{R}$ receives $k$ information bits from $\mathbb{S}$ and forwards them using another $n$ channel uses to the destination. Thus, here we consider that each block spans over $2n+v$ channel uses.
WPT and information transfer are carried out for every block, without any constraint on the minimum power level of the received signal, and all harvested energy\footnote{Herein, other consumption sources are neglected. In practice the energy used for transmission is usually much higher than the processing power required by the transceiver circuitry \cite{Nasir.2013,Nasir.2015,Chen.2014}.} at $\mathbb{R}$ is used when transmitting.
The duration of each channel use is denoted by $T_c$, while $T=(2n+v)T_c$ is the duration of the entire block. 
Also, perfect CSI at the receivers is assumed, as in \cite{Krikidis.2012,Chen.2014,Ding.2014,Nasir.2013,Nasir.2015}\footnote{The channel uses dedicated to obtaining CSI are not considered in our analysis. Then, the results obtained for a system where this quantity is not negligible are upper-bounded by our findings.}.

As in \cite{Nasir.2013}, we define the ratio between the portion of the block time used for WPT and the time used for information transfer as
\begin{equation}\label{alpha}
\alpha=\frac{vT_c}{(2n+v)T_c}=\frac{v}{2n+v}.
\end{equation}

\section{DF Relaying at Finite Blocklength}\label{DF}
The information transfer in DF relaying is affected by whether an error in decoding the signal occurs in the $\mathbb{S}\rightarrow\mathbb{R}$ channel or in the $\mathbb{R}\rightarrow\mathbb{D}$ channel. If an error occurs when $\mathbb{R}$ is decoding the message from the source, then $\mathbb{R}$ keeps inactive during the time reserved for its transmission, i.e. the last $n$ channel uses. 
This behavior reduces hardware complexity compared to other works, e.g. \cite{Nasir.2015} for infinite blocklength and preset power relay.

\subsection{Signal Model}

The received signal at $\mathbb{R}$ is given by
\begin{equation}\label{yrAF}
y_{r,i}=\frac{1}{\sqrt{d_1^{\omega}}}\sqrt{P_s}\tilde{h}_is_i+n_{r,i},
\end{equation}
where $i$ is the block index, $P_s$ is the $\mathbb{S}$ transmission power, $\omega$ is the path loss exponent, $n_{r,i}$ is the AWGN noise at $\mathbb{R}$ and $s_i$ is the normalized information signal from $\mathbb{S}$, i.e. $\mathbb{E}[|s_i|^2]=1$, where $\mathbb{E}[\mathrel{\;\cdot\;}]$ is the expectation operator.  Thus, the instantaneous Signal-to-Noise Ratio (SNR) at $\mathbb{R}$ is

\begin{equation}\label{SNR1}
\gamma_{_{r,i}}=\frac{P_s|\tilde{h}_i|^2}{d_1^{\omega}\sigma_r^2}=\frac{P_sh_{i}}{d_1^{\omega}\sigma_r^2},
\end{equation}
where $\sigma_r^2$ is the variance of the AWGN at $\mathbb{R}$. Then, the harvested energy during the time $\alpha T=vT_c$ is \cite{Nasir.2013}

\begin{equation}\label{Eh}
E_{h,i}=\frac{\eta P_s|\tilde{h}_i|^2}{d_1^{\omega}}\alpha T=\frac{\eta P_sh_{i}v}{d_1^{\omega}(2n+v)}T,
\end{equation}
where $0<\eta<1$ is the energy conversion efficiency which depends on the rectification process and the EH circuitry \cite{Zhou.2013}.

Let $L\in\{0,1,2,...\}$ be a random variable which accounts for the number of consecutive blocks being incorrectly decoded at $\mathbb{R}$ starting from the $i$-th block. Then, $y_{r,i+L}$ is correctly decoded and the received signal at $\mathbb{D}$ is
\begin{equation}\label{yd5}
y_{d,i+L}=\frac{1}{\sqrt{d_2^{\omega}}}\sqrt{P_{r,i+L}}\tilde{g}_{i+L}s_{i+L}+n_{d,i+L},
\end{equation}
where $n_{d}$ is the AWGN noise at $\mathbb{D}$. Notice that $P_{r,i+L}$ is the relay transmit power for the ($i+L$)-th block, which depends on the amount of harvested energy, and based on \eqref{Eh} it is

\begin{align}\label{Pr}
P_{r,i+L}&=\frac{E_{h,i+L}+E_{h,i+L-1}+...+E_{h,i}}{nT_c}\nonumber\\
&=\frac{1}{nT_c}\sum\limits_{l=0}^{L}E_{h,i+l}=\frac{\eta P_sv}{d_1^{\omega}n}\sum\limits_{l=0}^{L}h_{i+l}.
\end{align}
Also, the Probability Mass Function (PMF) of $L$ is given by
\begin{equation}\label{PDF}
P_L(l\!=\!z)\!=\!\left\{ \begin{array}{ll}
1\!-\!\epsilon_{r,i},&\! \mathrm{for}\ z=0 \\
(1\!-\!\epsilon_{r,i+z})\prod\limits_{j\!=\!0}^{z\!-\!1}\epsilon_{r,i+j},&\! \mathrm{for}\ z=1,2,...
\end{array},
\right.
\end{equation}
where $\epsilon_{r,i}$ is the error probability in decoding the $i$-th block at $\mathbb{R}$. Here we omit the proof since it is easy to check that $\sum_{z=0}^{\infty}P_L(l=z)=1$. Since $\epsilon_{r,i}$ is also a random variable depending on the $\mathbb{S}\rightarrow \mathbb{R}$ channel realization $h_i$, it seems intractable to deal directly with \eqref{Pr}. Taking into account that it is expected that $\epsilon_{r,i} \ll 1$ in URC scenarios, we can approximate \eqref{Pr} by setting $L=0$ as follows

\begin{align}\label{Pr2}
P_{r,i}\approx\frac{E_{h,i}}{nT_c}=\frac{\eta P_svh_{i}}{d_1^{\omega}n},
\end{align}
which is equivalent to not consider the energy harvested for those blocks that were incorrectly decoded at $\mathbb{R}$. From now on we work analytically using this approximation. Nevertheless, in Section~\ref{results} simulations are used in order to evaluate the performance when all the EH resources are taken into account, showing the accurateness of the proposed approximation. 

Using \eqref{Pr2} and \eqref{yd5} we can express the instantaneous SNR at $\mathbb{D}$ as follows

\begin{equation}
\label{SNR22}
\gamma_{_{d,i}}=\frac{P_{r,i}|\tilde{g}_i|^2}{d_2^{\omega}\sigma_d^2}=\frac{P_{r,i}g_i}{d_2^{\omega}\sigma_d^2} =\frac{\eta P_sh_ig_iv}{d_1^{\omega}d_2^{\omega}\sigma_d^2n},
\end{equation}
where $\sigma_d^2$ is the variance of the AWGN at $\mathbb{D}$.

\subsection{Performance at Finite Blocklength}
In the delay-limited transmission mode the throughput is determined by evaluating the error probability $\epsilon$ at a fixed transmission rate, $r=k/n$, in bits per channel use. For a single hop transmission, the error probability in a block of length $n$ and coding rate $r$ is given by \cite{Hu.2016},
\begin{equation}\label{e}
\epsilon=\mathbb{P}(\gamma,r,n)\approx Q\biggl(\frac{C(\gamma)-r}{\sqrt{V(\gamma)/n}}\biggl),
\end{equation}
where $C(\gamma)=\log_2(1+\gamma)$ is the Shannon capacity, $V(\gamma)=\big(1-\frac{1}{(\gamma+1)^2}\big)(\log_2e)^2$ is the channel dispersion, which measures the stochastic variability of the channel relative to a deterministic channel with the same capacity \cite{Polyanskiy.2010}, and $Q(z)=\int_{z}^{\infty}\frac{1}{\sqrt{2\pi}}e^{-t^2/2}dt$. Note that \eqref{e} is a valid approximation for sufficiently large values of $n$ \cite{Polyanskiy.2010, Yang.2014}, e.g. $n\ge100$. For quasi-static fading channels the error probability is \cite{Yang.2014}
\begin{equation}\label{pout}
\mathbb{E}[\;\epsilon\;]=\mathbb{E}\big[\mathbb{P}(\gamma,r,n)\big]\approx\mathbb{E}\Biggl[ Q\biggl(\frac{C(\gamma)-r}{\sqrt{V(\gamma)/n}}\biggl)\Biggl].
\end{equation}

Given that the transmitter is communicating with coding rate $r$ during a portion $n/(2n+v)$ of the block time, as shown in Fig.~\ref{fig:TSR}, the throughput or effective rate $\tau$ is then
\begin{equation}\label{tauDF}
\tau=(1-\mathbb{E}[\epsilon_{_{\mathrm{DF}}}])\frac{k}{2n+v},
\end{equation}
where $\epsilon_{_{\mathrm{DF}}}$ is the block error probability which depends on the $\mathbb{S}\rightarrow\mathbb{R}$ error probability ($\epsilon_r$), and $\mathbb{R\rightarrow\mathbb{D}}$ error probability ($\epsilon_d$). This dependency is given by
\begin{equation}\label{eDF}
\epsilon_{_{\mathrm{DF}}}=\epsilon_{_r}+(1-\epsilon_{_r})\epsilon_{_d}=\epsilon_{_r}+\epsilon_{_d}-\epsilon_{_r}\epsilon_{_d},
\end{equation}
\begin{equation}\label{eDF2}
\mathbb{E}[\epsilon_{_{\mathrm{DF}}}]=\mathbb{E}[\epsilon_{_r}]+\mathbb{E}[\epsilon_{_d}]-\mathbb{E}[\epsilon_{_r}\epsilon_{_d}],
\end{equation}
where $\epsilon_{_r}=\mathbb{P}(\gamma_{_r},r,n)$ and $\epsilon_{_d}=\mathbb{P}(\gamma_{_d},r,n)$ are given by \eqref{e} for each channel realization. Note that $\mathbb{E}[\epsilon_{_r}\epsilon_{_d}]\ne \mathbb{E}[\epsilon_{_r}]\mathbb{E}[\epsilon_{_d}]$ since $\epsilon_{_r}$ and $\epsilon_{_d}$ depend on $\gamma_{_r}$ and $\gamma_{_d}$, respectively, which are correlated
\footnote{Notice that without EH, $\mathbb{E}[\epsilon_{_{\mathrm{r}}}\epsilon_{_{\mathrm{d}}}]= \mathbb{E}[\epsilon_{_{\mathrm{r}}}]\mathbb{E}[\epsilon_{_{\mathrm{d}}}]$ as in \cite{Hu.2016}.
}, i.e. both depend on $\mathbb{S}\rightarrow\mathbb{R}$ channel realization $\tilde{h}$. Each term in \eqref{eDF2} can be expressed using \eqref{pout} as follows

\begin{align}
\mathbb{E}[\epsilon_{_r}]&=\mathbb{P}\big[\gamma_{_r},r,n\big] \nonumber\\
&\approx\mathbb{E}\Biggl[Q\biggl(\frac{C(\gamma_{_r})-r}{\sqrt{V(\gamma_{_r})/n}}\biggl)\Biggl] \nonumber\\
&\approx\int\limits_{0}^{\infty}f_h(h|m)\epsilon_r\mathrm{d}h\approx\frac{m^m}{\Gamma(m)}\int\limits_{0}^{\infty}h^{m-1}e^{-mh}\epsilon_{_r}\mathrm{d} h \nonumber\\
\label{er}
&\approx\frac{m^m}{\sqrt{2\pi}\Gamma(m)}\int_{0}^{\infty}\int_{z(\gamma_{_r})}^{\infty}h^{m-1} \exp\left(-\frac{t^2}{2}-mh\right)\mathrm{d}t\mathrm{d}h,\\
\mathbb{E}[\epsilon_{_d}]&=\mathbb{E}\big[\mathbb{P}(\gamma_{_d},r,n)\big]\nonumber\\
&\approx\mathbb{E}\Biggl[Q\biggl(\frac{C(\gamma_{_d})-r}{\sqrt{V(\gamma_{_d})/n}}\biggl)\Biggl] \nonumber\\
&\approx\int_{0}^{\infty}\int_{0}^{\infty}f_h(h|m)f_g(g|m)\epsilon_{_d}\mathrm{d}h\mathrm{d}g\nonumber\\
\label{ed}
&\approx\int\limits_{0}^{\infty}\int\limits_{0}^{\infty}\dfrac{m^mh^{m-1}e^{-mh}}{\Gamma(m)}\frac{m^mg^{m-1}e^{-mg}}{\Gamma(m)}\epsilon_{_d}\mathrm{d}h\mathrm{d}g\nonumber\\
&\approx\frac{m^{2m}}{\sqrt{2\pi}\Gamma(m)^2}\!\int\limits_{0}^{\infty}\!\int\limits_{0}^{\infty}\!\int\limits_{z(\gamma_{_d})}^{\infty}\!(hg)^{m\!-\!1} \exp\left(-\frac{t^2}{2} - m(h\!+\!g) \right)\mathrm{d}t\mathrm{d}h\mathrm{d}g,
\end{align}

\begin{align}\label{ered}
\mathbb{E}[\epsilon_{_r}\epsilon_{_d}]&=\mathbb{E}\big[\mathbb{P}(\gamma_{_r},r,n)\mathbb{P}(\gamma_{_d},r,n)\big]\nonumber\\
&\approx\mathbb{E}\Biggl[Q\biggl(\frac{C(\gamma_{_r})-r}{\sqrt{V(\gamma_{_r})/n}}\biggl)Q\biggl(\frac{C(\gamma_{_d})-r}{\sqrt{V(\gamma_{_d})/n}}\biggl)\Biggl]\ \ \ \ \ \ \nonumber\\
&\approx\int\limits_{0}^{\infty}\int\limits_{0}^{\infty}f_h(h|m)f_g(g|m)\epsilon_{_r}\epsilon_{_d}\mathrm{d}h\mathrm{d}g \nonumber\\
&= \frac{m^{2m}}{2\pi\Gamma(m)^2}
\int\limits_{0}^{\infty}\int\limits_{0}^{\infty}\int\limits_{z(\gamma_{_r})}^{\infty}\int\limits_{z(\gamma_{_d})}^{\infty}\!(hg)^{m\!-\!1} \exp\left(-\frac{t_1^2\!+\!t_2^2}{2}-m(h\!+\!g)\right) \mathrm{d}t_1\mathrm{d}t_2 \mathrm{d}h\mathrm{d}g,
\end{align}
where $z(\gamma_{_r})$ and $z(\gamma_{_d})$ are given by $z(\gamma)=\tfrac{C(\gamma)-r}{\sqrt{V(\gamma)/n}}$. Then, the corresponding average delay (measured in number of channel uses) is
\begin{equation}\label{delay}
\delta=\frac{k}{\tau}=\frac{2n+v}{1-\mathbb{E}[\epsilon_{_{\mathrm{DF}}}]}.
\end{equation}

Notice that finding closed-form solutions for \eqref{er}, \eqref{ed} and \eqref{ered} also seems to be intractable. Therefore, next we resort to  numerical integration in order to evaluate the system performance.

\section{Numerical Results}\label{results}
Next we present numerical results regarding the performance of the TSR protocol with DF relaying at finite blocklength. We use numerical evaluation in order to characterize the throughput and delay in constrained error scenarios with latency requirements, which will be typical when considering URC services in 5G systems. Thus, let $\epsilon_{_0}$ be the target error probability where $\epsilon_{_{\mathrm{DF}}}\le\epsilon_{_0}$ must be satisfied. Simulations are also used in order to check the real performance of the DF relaying scheme taking into account the energy stored when $\mathbb{R}$ fails in decoding the signal from $\mathbb{S}$, thus using \eqref{Pr} instead of \eqref{Pr2}.

The distances $d_1$ and $d_2$ are normalized to unit value and the shape factor $m$ is chosen to be 2, which means that the links experience multi-path as well as some line-of-sight. 
All results are obtained by setting the $\mathbb{S}$ transmission power to $P_s=1$ Joule/sec and the path loss exponent $\omega=2.7$ as in \cite{Nasir.2013}. As the state-of-the-art in circuit design establishes that RF signals over a wide range of frequencies can be rectified at an efficiency higher than 50\% \cite{Lu.2015}, we consider $\eta=0.5$. For simplicity, equal noise variances at $\mathbb{R}$ and $\mathbb{D}$ are assumed, $\sigma_r^2=\sigma_d^2=0.01$. 
\begin{figure}[ht!]
	\centering
	\subfigure{\label{fig:r1a}\includegraphics[width=0.8\textwidth]{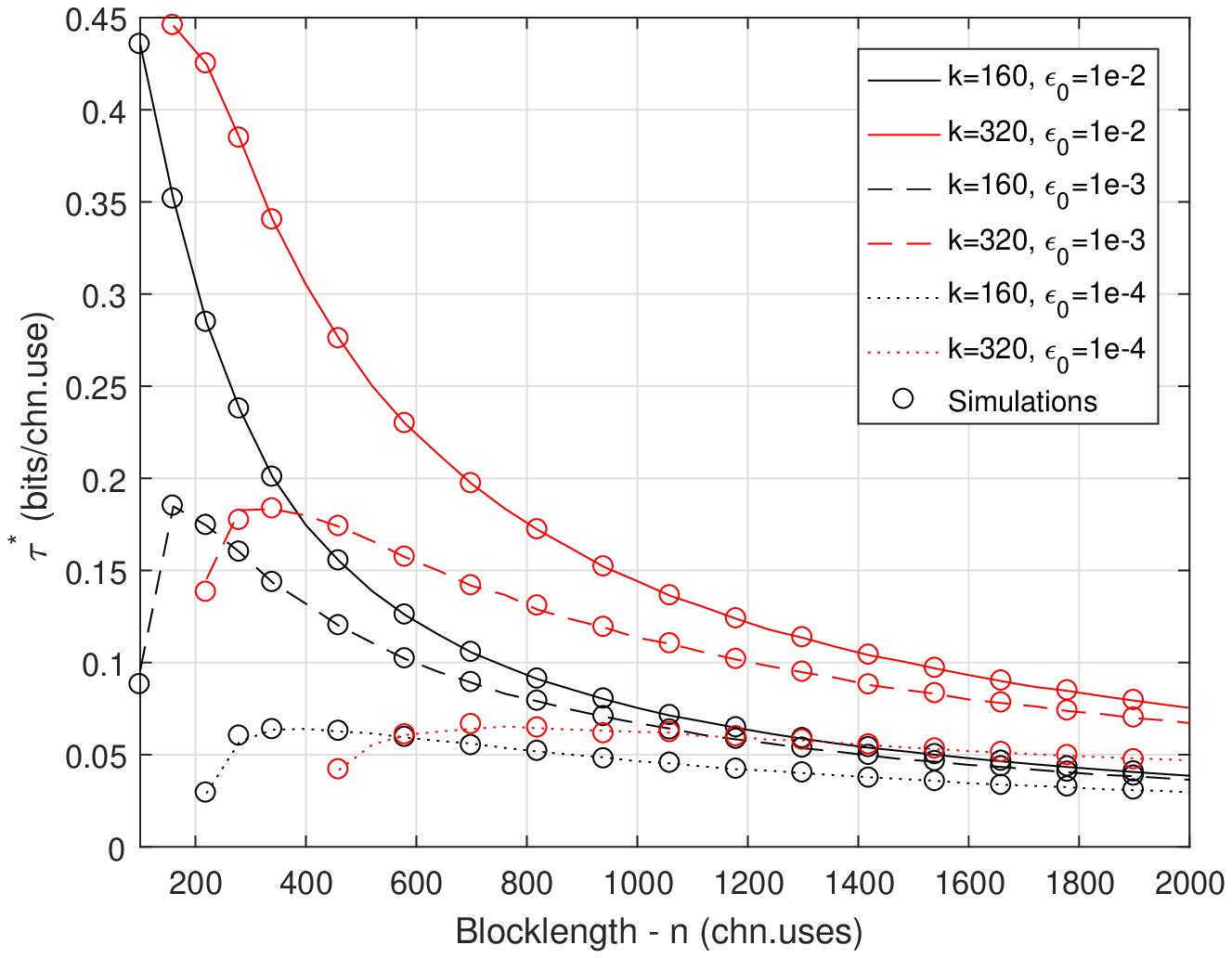}}
	\subfigure{\label{fig:r1b}\includegraphics[width=0.8\textwidth]{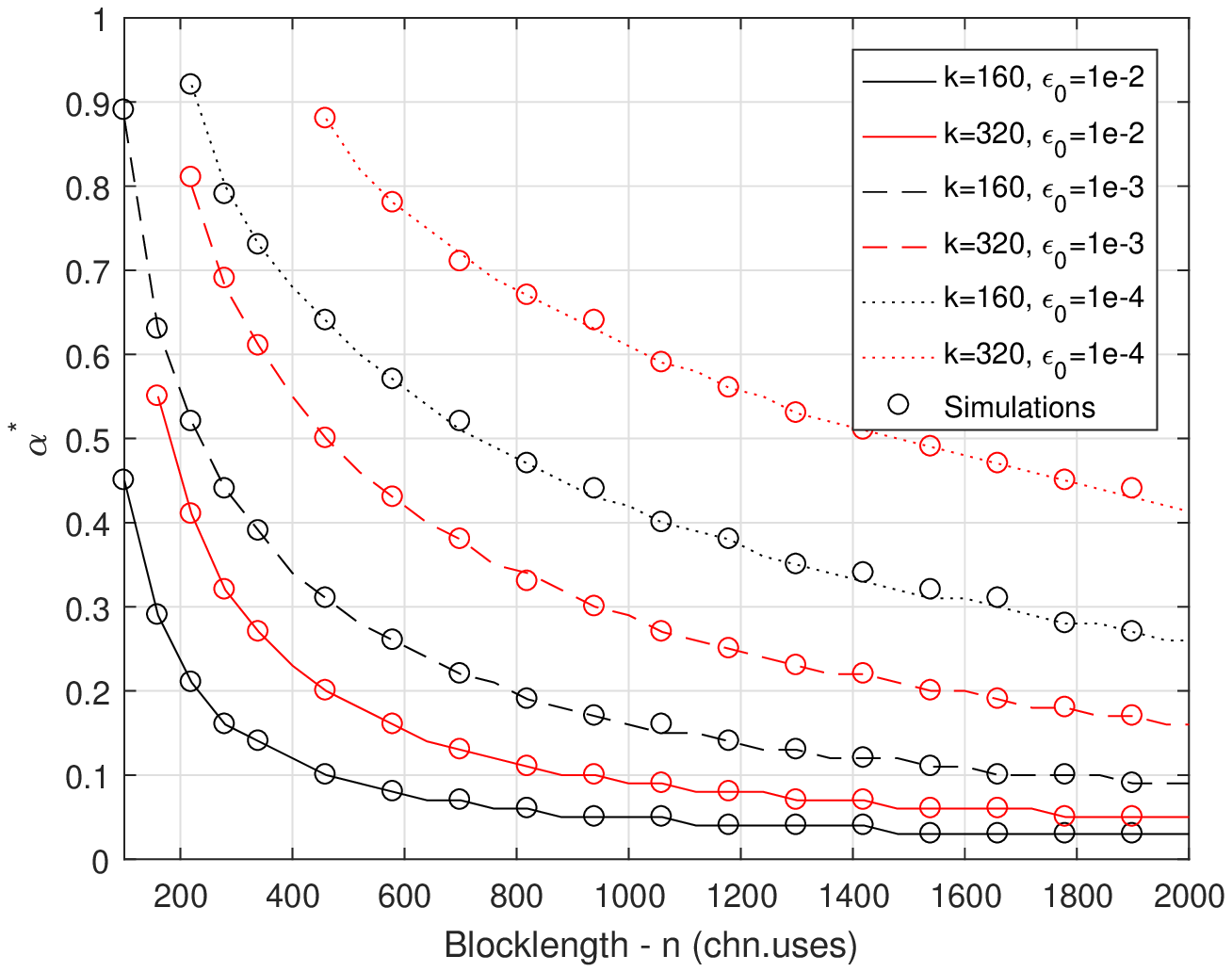}}	
	\vspace{-2mm}
	\caption{(a) $\tau^*$ (top) and (b) $\alpha^*$ (bottom), for $k=160$ and $k=320$ bits, as a function of blocklength.}\label{fig:r1}		
\end{figure}

Fig.~\ref{fig:r1} shows the performance when messages of $k=160$ and $k=320$ bits are transmitted over $100\le n\le 2000$ channel uses per hop. Particularly, Fig.~\ref{fig:r1}a (top) shows the maximum throughput ($\tau^*$) in bits per channel use for the optimal value of $\alpha$, ($\alpha^*$), which is plotted in Fig.~\ref{fig:r1}b (bottom). First, we can note that numerical evaluation considering \eqref{Pr2} agrees very well with simulations using \eqref{Pr} for $\epsilon_0\ll 1$, which validates the approximation made in \eqref{Pr2}. Second, there is an optimum blocklength, $n^*$, which increases as $\epsilon_0$ decreases and, at the same time, the optimum required proportion of time reserved for PT ($\alpha^*$) increases as well. Note that for $\epsilon_0=10^{-2}\rightarrow n^*<100$. Third, increasing $k$ has a similar effect, and the same optimum throughput is attainable independently from $k$ but requiring different values of $n$ and $v$. 
\begin{figure}[ht!]
	\centering
	\subfigure{\label{fig:r1a}\includegraphics[width=0.8\textwidth]{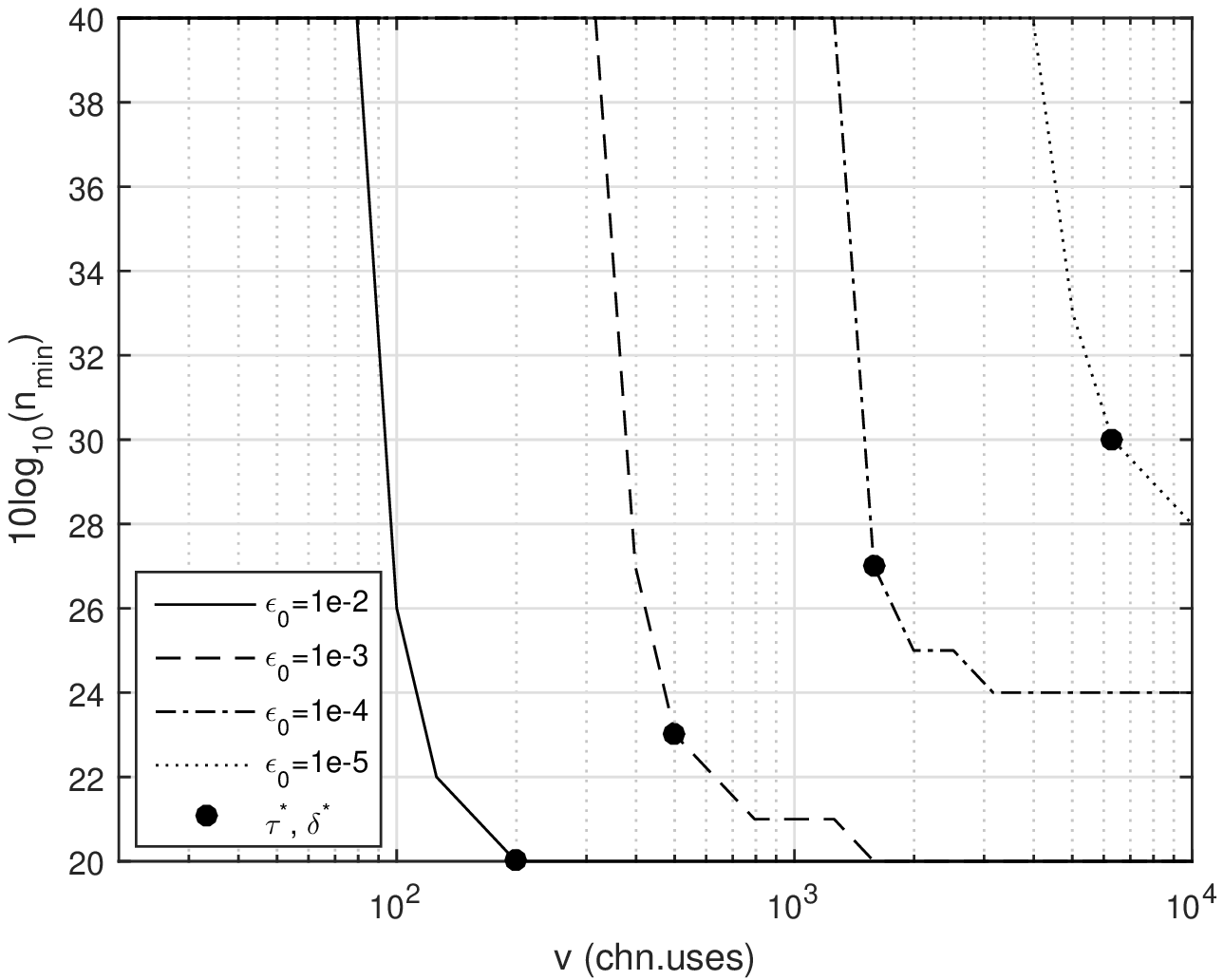}}
	\vspace{-4mm}
	\caption{Minimum $n$ as a function of $v$.}		
	\label{fig_r2}
\end{figure}

\begin{figure}[t!]
	\centering
	\subfigure{\label{fig:a}\includegraphics[width=0.8\textwidth]{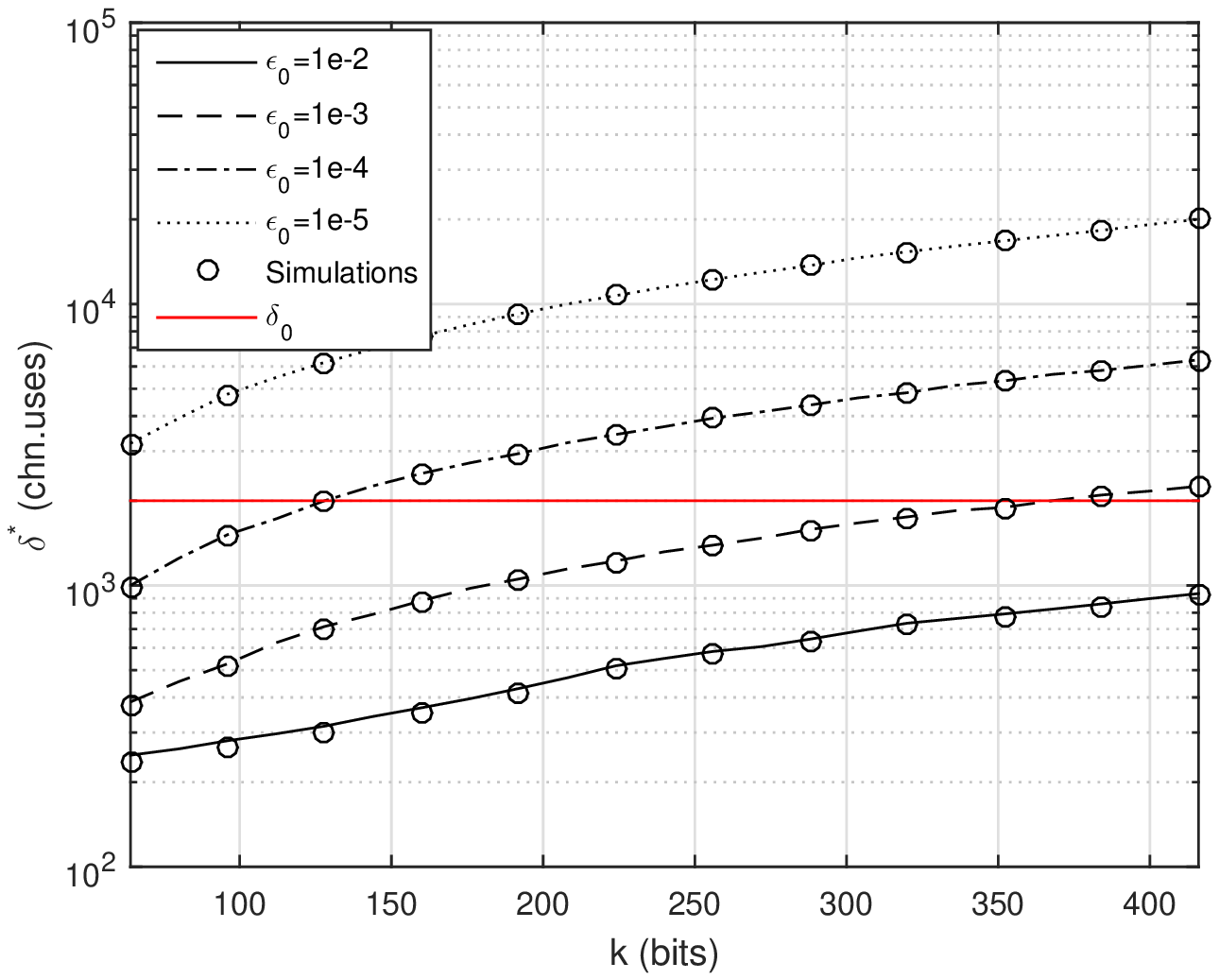}}
	\subfigure{\label{fig:b}\includegraphics[width=0.8\textwidth]{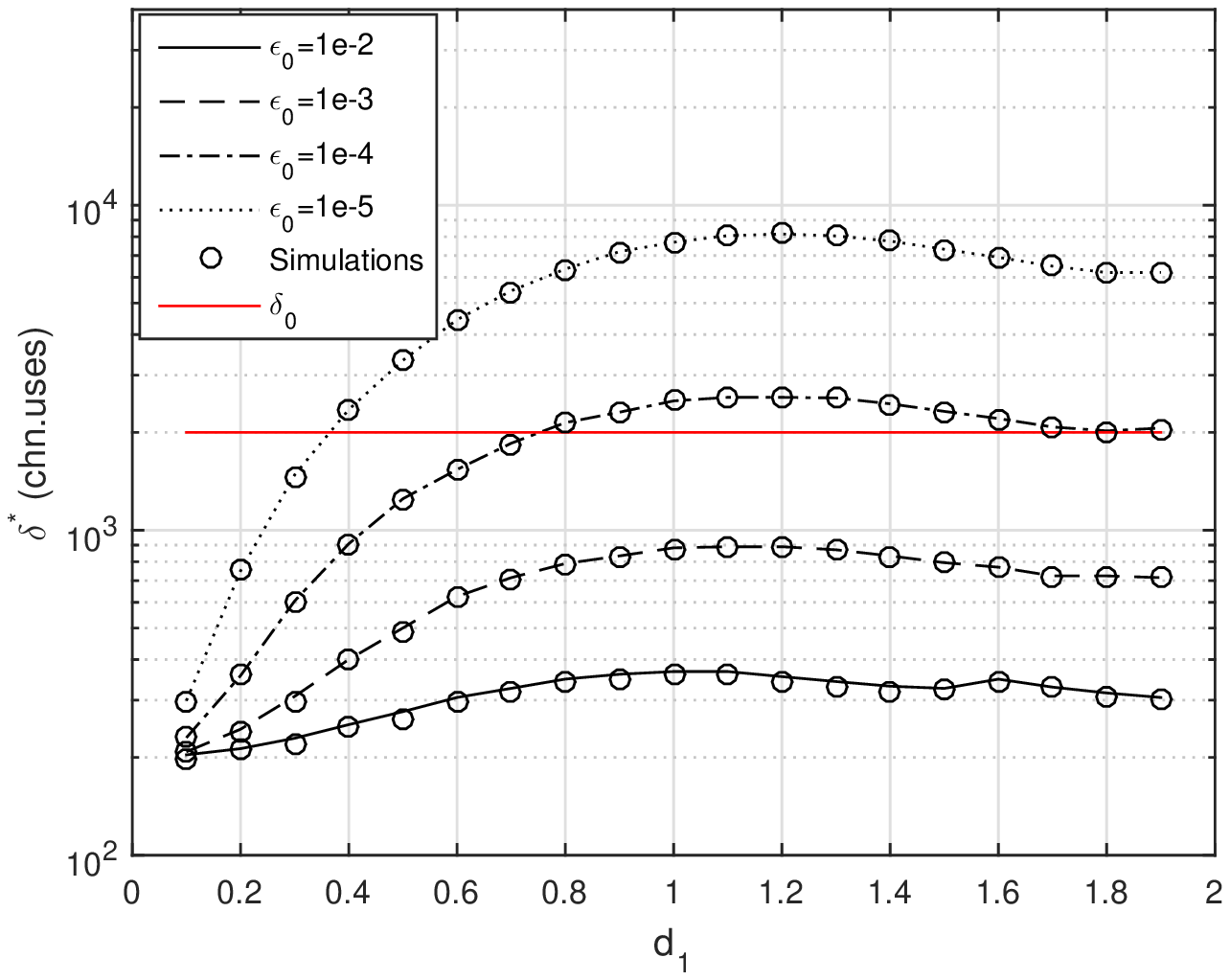}}	
	\vspace{-4mm}
	\caption{Minimum delay ($\delta^*$) in channel uses, as a function of: (a) packet lengths ($k$) in bits (top), (b) relay position ($d_1$, with $d_2=2-d_1$) (bottom).}	
	\label{fig_r3}	
\end{figure}

The minimum required number of channel uses for successful information transfer ($n_{\mathrm{min}}$) is plotted in Fig.~\ref{fig_r2} as a function of the number of channel uses for PT ($v$), for different reliability requirements and messages of $k=160$ bits. When the time for PT increases, more energy is harvested and the $\mathbb{R}$ transmit power increases as well, so a smaller number of channel uses ($n$) for transmission is required.  As expected, the required values for $v$ and $n$ are greater when the required reliability increases, which impacts on the achievable throughput and delay since they depend strongly on $2n+v$. The points for maximum throughput ($\tau^*$) and minimum delay ($\delta^*$) are also marked in Fig.~\ref{fig_r2}. Notice that for $\epsilon_0=10^{-5}\rightarrow v^*\approx6000$ channel uses and $n^*=1000$, then $\delta^*\approx\frac{2n^*+v^*}{1-10^{-5}}\approx8001$ channel uses, which could be very severe depending on the value of $T_c$, e.g. for $T_c=2\mu s\rightarrow \delta^*\approx 16$m$s$. 

The size of the message ($k$) is a parameter that strongly impacts on the required number of channel uses, i.e. $n$ and $v$, to achieve certain throughput, as we previously showed in Fig.~\ref{fig:r1}a. This fact is further discussed through Fig.~\ref{fig_r3}a, where we plot the minimum delay as a function of $k$, for scenarios with different reliability requirements. Notice that $\delta^*=k/\tau^*$, according to \eqref{delay}. When stringent reliability is mandatory, e.g. $\epsilon_0=10^{-5}$, the delay could reach very large values, e.g. $\delta^*>3000$ channel uses when $k>64$ bits. For illustrative purposes, we delimit with a red line in Fig.~\ref{fig_r3} a given maximum allowed delay ($\delta_0=2000$ channel uses). Under $\delta_0$ and $\epsilon_0$ requirements the possible values of $k$ are limited. In that sense, when $\epsilon_0=10^{-3}\rightarrow k<350$ bits, and when  $\epsilon_0=10^{-4}\rightarrow k<130$ bits, while messages which need to be transmitted with $\epsilon_0=10^{-5}$ must carry a very low number of information bits.

All previous results considered $\mathbb{R}$ just mid-way between $\mathbb{S}$ and $\mathbb{D}$. In Fig.~\ref{fig_r3}b, $\delta^*$ is plotted as a function of the relay position ($d_1$, while setting $d_2=2-d_1$) for messages of $k=160$ bits. The minimum delay (corresponding to the maximum throughput) is achieved when $\mathbb{R}$ is closer to $\mathbb{S}$. This is in agreement with previous results for this scenario at infinite blocklength and AF relaying \cite{Nasir.2013}, and in contrast to the scenario where WPT is not considered \cite[Chapter~16]{Liu.2009}. Now, even very stringent reliability requirements, such as  $\epsilon_0=10^{-4}$ and $\epsilon_0=10^{-5}$, can be met when $d_1\le0.65$ and $d_1\le0.35$, respectively. From both Fig.~\ref{fig_r3}a and Fig.~\ref{fig_r3}b we note that simulation results agree very well again with analytical approximations.
\begin{figure}[h!]
	\centering
	\subfigure{\label{fig:r1a}\includegraphics[width=0.8\textwidth]{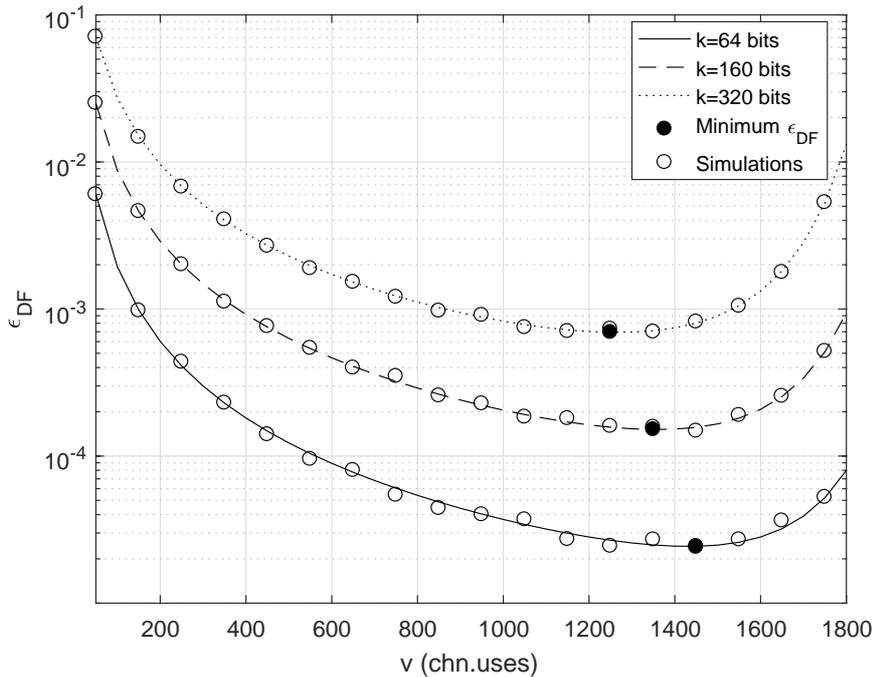}}
	\vspace{-4mm}
	\caption{Error probability ($\epsilon_{_{\mathrm{DF}}}$) as a function of $v$ for $2n+v=2000$ channel uses.}		
	\label{fig_r4}
\end{figure}

In Fig.~\ref{fig_r4}, we set $2n+v=2000$ channel uses and plot the error probability as a function of the number of channel uses for PT ($v$) for messages with 64, 160 and 320 information bits. Then, from \eqref{delay} we have that $\delta\approx2000$ channel uses as long as $\epsilon_{_{\mathrm{DF}}}\ll 1$, which holds for practical URC scenarios. As shown in this figure, there is a trade-off between $n$ and $v$, and the existence of an optimal point can be explained as follows. Note that increasing $v$ at the expense of $n$ improves $\gamma_{_{d}}$, which favors $C(\gamma)$ in \eqref{e}. On the other hand, decreasing $n$ can affect negatively the error probability in \eqref{e} because both the rate $r=k/n$ and the term $\sqrt{n}$ increase.  Also, now $\alpha=v/2000$, and differently from Fig.~\ref{fig:r1}b for an unlimited delay, here $\alpha^*$ increases slowly when $k$ decreases\footnote{Notice that here $\alpha^*$ is the value of $\alpha$ for which $\epsilon_{_{\mathrm{DF}}}$ is minimum, which matches with $\tau^*$ since $2n+v$ is fixed.}. This is because when $k$ decreases, the fixed rate reduces its impact on the error probability. Finally, we can see that very short packets, e.g. with $k=64$ bits, can be transmitted ultra-reliably, e.g. with $\epsilon_0\approx 3\times 10^{-5}$, with acceptable latency, e.g. $\delta\approx2000$ channel uses,  using $v\approx1500$ and $n\approx500$ channel uses.

\section{Conclusion}\label{conclusions}
In this paper we evaluated a dual-hop DF relaying scheme for the delay-limited transmission mode at finite blocklength, where the relay is energy constrained and operates according to the TSR protocol \cite{Nasir.2013} over Nakagami-m channels. We derived an analytical approximation for the throughput of URC with DF relaying in such scenario and validated its accuracy through simulations. The numerical results show that, for URC scenarios with reliability requirements and latency constraints, a trade-off is posed. While reliability requires increasing the number of channel uses for PT and/or information transfer, such increase affects negatively the latency. Moreover, we show that in general the possibility of meeting the reliability and latency constraints increases by decreasing the message length and/or for relay positions closer to the source. Finally, performance analysis under different target error probabilities and latency constraints are discussed, where we show that operation under typical URC scenarios is feasible. 

\section*{Acknowledgment}
This work was supported by CNPq, CAPES, Funda\c{c}\~ao Arauc\'aria (Brazil) and Academy of Finland, including a scholarship from the Program for Graduate Students from Cooperation Agreements (PEC-PG, of CAPES/CNPq Brazil).
\bibliographystyle{IEEEtran}
\bibliography{IEEEabrv,references}
\end{document}